\documentclass[showkeys]{revtex4-1}

\usepackage{amssymb}
\usepackage{amsfonts} 

\usepackage{amsthm} 
\usepackage{amsmath} 
\usepackage{graphicx}

\graphicspath{ {Figures/} }

\begin{document}

\title{Modeling COVID-19 pandemic with financial markets models: The case of Jaén (Spain)}

\author{Julio Guerrero}
\email{jguerrer@ujaen.es}
\affiliation{Department of Mathematics, University of Ja\'en, Campus Las Lagunillas s/n, 23071 Ja\'en, Spain}
\affiliation{Institute Carlos I of Theoretical and Computational Physics (iC1), University of  Granada,
Fuentenueva s/n, 18071 Granada, Spain}

\author{Maria del Carmen Galiano}
\affiliation{Department of Mathematics, University of Ja\'en, Campus Las Lagunillas s/n, 23071 Ja\'en, Spain}

\author{Giuseppe Orlando}
\email{giuseppe.orlando@uniba.it: corresponding author}
\affiliation{Department of~Economics and Finance, University of Bari, Largo Abbazia S.~Scolastica, Bari, 70124, Italy}

\begin{abstract}
The main objective of this work is to test whether some stochastic models typically used in financial markets could be applied to the COVID-19 pandemic. To this end we have implemented the ARIMAX and Cox-Ingersoll-Ross (CIR) models originally designed for interest rate pricing but transformed by us into a forecasting tool. For the latter, which we denoted CIR*, both the Euler-Maruyama method and the Milstein method were used. Forecasts obtained with the maximum likelihood method have been validated with 95\% confidence intervals and with statistical measures of goodness of fit, such as the root mean square error (RMSE). We demonstrate that the accuracy of the obtained results is consistent with the observations and sufficiently accurate to the point that the proposed CIR* framework could be considered a valid alternative to the classical ARIMAX for modelling pandemics. 
\end{abstract}


\maketitle

\section{Introduction}

Coronavirus disease 2019 (COVID-19) is a lung disease caused by severe acute respiratory syndrome coronavirus 2 (SARS-CoV-2). In December 2019, the Chinese authorities reported different cases of this virus in Wuhan. This disease spread rapidly throughout the world from less than 30 cases at the end of December 2019 to more than 8,455,738 confirmed cases on June 20, 2021.

The first case in Spain was a German tourist, on January 31, 2021. From that moment, several cases began to be confirmed throughout the country. In Andalusia, the first positive was detected in Seville on February 26, 2020. Two days later the first case was confirmed in the province of Jaén, and on March 6 the first case of coronavirus in the city of Jaén.

As the days went by, after the high number of infections in the country, the state of alarm was decreed on March 14, which was extended until June 21.
The application of measures such as the use of a mask, perimeter confinements, the closure of non-essential services, etc., improved the infection rate two weeks after the declaration of quarantine.
During these years, many researchers from various disciplines have used various modelling tools to analyze the impact of the pandemic at the global and local levels. In our case, we are going to focus on modelling the contagion in Jaén in two different ways.
The first approach is based on the Autoregressive Integrated Moving Average with Explanatory Variable (ARIMAX) model \cite{Mills1990} which, we found, provides better performances than the  Autoregressive Integrated Moving Average (ARIMA) (on the same line, see also \cite{lee2010calendar, Chadsuthi2012Jul, Suharsono2015Dec,Anggraeni2015Jan, Shilpa2019}).
This is a common model in time series forecasting and is often adopted in finance \cite{Ariyo2014, Subramaniam2020, Orlando2022Jun, Orlando2022Feb}.  

The second approach is based on the Cox, Ingersoll and Ross (CIR) model. This is a model designed for interest rates pricing that we turn into a forecasting tool. We prove that this transformation of the CIR model,  which we have denoted CIR*, outperforms the classical ARIMAX, For the latter, both the Euler-Maruyama method and the Milstein method were used and this is the main contribution of the present study. 
Notice that the suggested approach not only extends the models available to scholars to model pandemics but, also, paves the way for similar approaches where financial models can be converted into econometric models.

For the implementation of real-world data, we use the data from the moving averages for 14 days of the daily cases of city of Jaén, that is, each data represents the average number of people infected in each of the previous 14 days. This is because, due to the weekend effect and occasional misreporting, we found that the moving average is a more reliable target. This is due to two reasons: a) the relatively small size of the city of Jaén, which  affects the number of cases and b) the effect of the weekend when reporting is altered. The two effects lead to highly irregular behavior in the time series considered.
Over that time series, we estimate the parameters of the considered models, in such a way that they best fit the data. The forecasts have been validated with 95\% confidence intervals and with statistical measures of goodness of fit, such as the RMSE.

This article is organized as follows. Section \ref{LitRev} briefly summarizes the literature. Section \ref{MatMeth} reports the data,  describes the CIR model as well as the methodology on which it is based. That is followed by the explanation the suggested adaptation to forecasting, by its calibration and by a sketch of the ARIMAX model. 
Section \ref{Result} shows the obtained results of the two models by comparing them.  Section \ref{Conclusion} concludes.

\section{Literature review}
\label{LitRev}

Among those works that adopted the ARIMA (and the like models) to estimate the cases of the COVID-19 pandemic, we mention Ekinci \cite{Ekinci2021} who considered data from  USA, India, Brazil, France, Russia, UK, Italy, Spain and Germany. When comparing ARMA-GARCH, ARMA-TGARCH and ARMA-EGARCH models, it was found that while considering the conditional variance effect improves the forecasting power, the asymmetric effect (such as asymmetric GARCH models) has mixed results. 
Sahai et al. \cite{Sahai2020} adopted the ARIMA for analyzing the trend of COVID-19 cases in Spain, Italy, France, Germany and the US. The authors claim that their model provides considerable forecast accuracy and could be a useful tool for governments to ramp up their healthcare preparations. 
Subramaniam et al. \cite{Subramaniam2020} draw a parallel between forecasting stock prices and cases of the pandemic by means of the ARIMA model. Then, they explore the correlation between the predictive efficiency of the ARIMA model and variation in the data.
Katoch et al. \cite{Katoch2021} adopt an ARIMA model to analyse the temporal dynamics of the COVID-19 outbreak in India from 30 January 2020 to 16 September 2020. Their approach suggests "varying epidemic’s inflexion point and final size for underlying states and the mainland, India".
Regarding the alternative between ARIMA and ARIMAX, in the literature has been found that the latter may yield  better forecast compared to the seasonal ARIMA (SARIMA) model and Neural Networks (e.g. see Suhartono \cite{Suhartono2015}). This is because the ARIMAX is most suited to deal with calendar effects \cite{lee2010calendar, Chadsuthi2012Jul, Suharsono2015Dec,Anggraeni2015Jan, Tanyavutti2018}. 

As regards the Cox, Ingersoll and Ross (CIR) model \cite{Cox1985,Cox2005Jul}, as already mentioned, it has been proposed for the pricing of interest rates. At the time of its introduction, it quickly gained popularity in finance because it was
perceived as "an improvement on the Vasicek model \cite{Vasicek1977}, not allowing for negative rates and
introducing rate-dependent volatility, as well as for its relatively handy implementation
and analytical tractability" \cite{orlando2018new}. 

Other applications of the CIR model include stochastic volatility modelling in option
pricing problems \cite{canale2017analytic, orlando2017review}  or default intensities in
credit risk \cite{Duffie2005credit, Orlando2021FinEng}.
In this study, similar to what has been done by Orlando and at. \cite{Orlando2019, orlando2018new, Orlando2019, orlando2020forecasting, Orlando2021Dec},
when developing the CIR\# model, we transform the original CIR model into a forecasting tool and compare its performance with that of the well known ARIMAX model.

\section{Materials and methods} \label{MatMeth}

\subsection{Data}
\begin{figure}[ht]
    \centerline{\scalebox{0.3}{
    \includegraphics{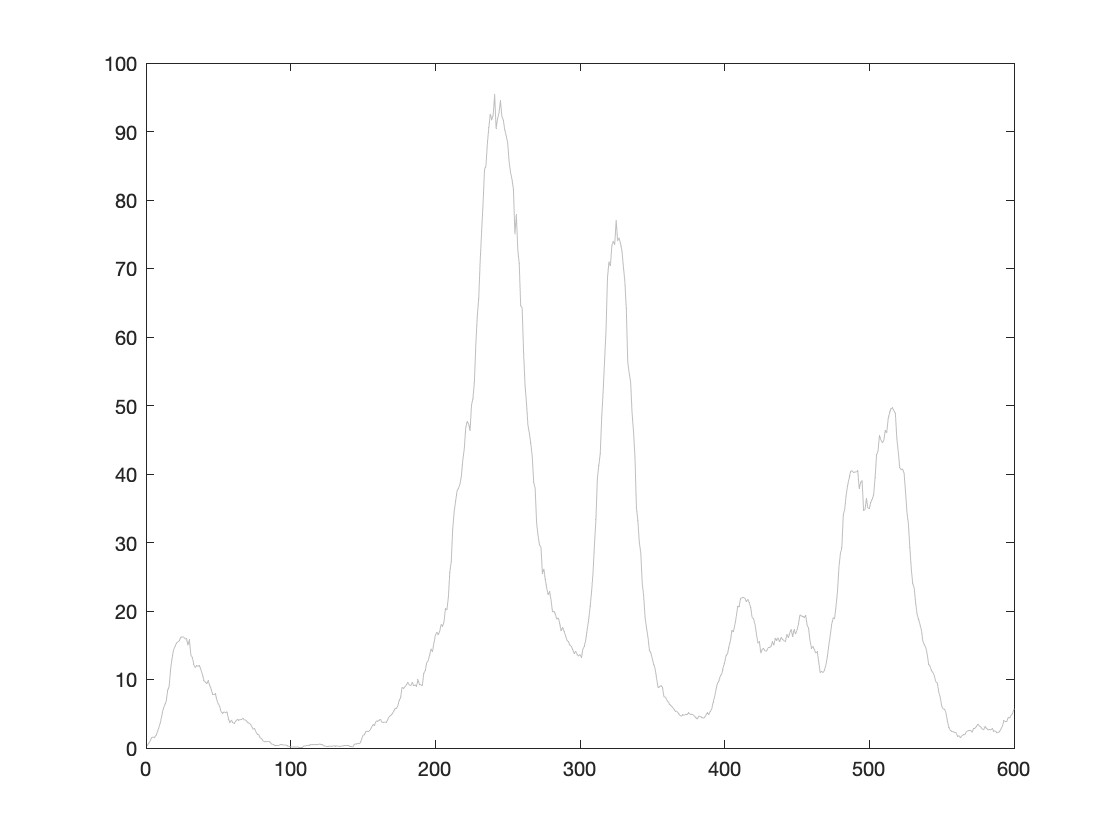}
    }}
    \caption{14-day moving average of the daily cases of COVID-19 in the city of Jaén. The abscissa represents the days elapsed since February 2, 2021.}
    \label{fig:23}
\end{figure}

The available data are the moving averages of 14 days of infections in the city of Jaén, from February 2, 2020 to October 8, 2021 and have been provided by the  Health and Family Council of the Andalusian Regional Government  (see Figure \ref{fig:23}). 
Jaén is a relatively small city (around 110 thousand inhabitants), which represents a challenge since the number of  COVID-19 daily cases  is small and  with large fluctuations, even after the moving averages are computed. Therefore we shall have the opportunity of testing different methods in unfavourable circumstances.\\

\subsection{ARIMAX model}

Leaving aside the cases where data show evidence of non-stationarity where an initial differencing removes the integrated (I) part,
%
the ARMAX model can be described as:
\begin{equation}
\begin{split}
y(t)+a_1 y(t-1)+ \ldots +a_{n_a} y(t-n_a)&=\\
 & b_1 x(t-n_k)+ \ldots +b_{n_b} x (t-n_k-n_b+1) +     \\
 & c_1 \varepsilon(t-1) + \ldots +c_{n_c} \varepsilon(t-n_c)+\epsilon(t)
\end{split}
\end{equation}
with, $y(t)$ dependent/output variable at time $t$, $n_a$, number of poles, $n_b$ number of zeroes plus 1, $n_c$ number of $c$ coefficients, $n_k$ dead time in the system. 
Moreover, $ y(t-1) \ldots y(t-n_a) $ denotes the dependence between the current output and the previous outputs, 
$x(t-n_k) \ldots x(t-n_k-n_b+1)$ indicates the dependence between the current output and both the previous and delayed inputs, and
$\epsilon(t)$ expresses a white-noise error.

The orders of the ARMAX model are given by the parameters $n_a, n_b$, and $n_c$ whilst $n_k$ is the delay and $q$ is the delay operator. 
The ARMAX in compact form can be written as

\begin{equation}
A(q) y(t) = B(q) x(t-n_k)+C(q) \varepsilon(t)    
\end{equation}

such that, 
\begin{equation*}
    \begin{split}
 A(q)&=1+a_1 q^{-1}+ \ldots +a_{n_a} q^{-n_a} \\
B(q)& =b_1+ b_2 q^{-1} + \ldots +b_{n_b} q^{-n_b+1} \\
C(q)& =1+c_1 q^{-1}+ \ldots +c_{n_c} q^{-n_c} .
    \end{split}
\end{equation*}

The ARIMAX model can be seen as a generalization of the ARIMA because adds to the structure above described an integrator in the white noise $\varepsilon(t)$ as follows:

\begin{equation}
A(q) y(t) = B(q)x(t-nk)+ \dfrac{C(q)}{(1-q^{-1})}  \varepsilon(t) .
\end{equation}


\subsubsection{Estimation of the ARIMAX model}
To estimate the ARIMAX model the following steps have been performed. 

\begin{enumerate}
    \item 
    Ensure stationarity of the times series by conducting Augmented Dicky Fuller  (ADF) Test.
    \item
    Model identification, i.e. specification of the autoregressive (AR) and moving average (MA) terms with the help of the autocorrelation function (ACF) and partial autocorrelation function (PACF).
    \item
    Parameter estimation according to Ljung \cite{Ljung1998} and related implementation in Matlab \cite{ARIMAXMatlab}. 
    The best model is selected based on Akaike information criterion (AIC) values \cite{Stoica2004}.
\end{enumerate}

From now on, we refer to the ARMAX model (and not to the ARIMAX) assuming that the integration (I) has been removed.

\subsection{CIR* model}

As mentioned, this model emerged in 1985 from the hand of John C. Cox, Jonathan E. Ingersoll and Stephen A. Ross \cite{Cox1985,Cox2005Jul} as an improvement of the Vasicek model to prevent negative interest rates. 

The CIR model is based on the following equation:

\begin{equation}\label{equ:6}
\left\lbrace
\begin{array}{ll}
dX_t=\alpha \left(\mu-X_t\right)dt+\sigma \sqrt{X_t} dW_t\\
X_0=x_0
\end{array}
\right.
\end{equation}

Here, $\alpha,\mu$ and $\sigma$ are positive constants, $X(t)$ is the interest rate, $t$ is time, and $W_t$ denotes the standard Wiener process.

The parameters include the following:

\begin{itemize}
    \item 
    $\alpha (\mu-X_t)$ is the same factor as in Vasicek's model, so the interpretation of the deterministic solution is the same.
    \item 
    The standard deviation factor $\sigma \sqrt{X_t}$ removes negative rates.
    \item 
    $\sqrt{X_t}$ increases the standard deviation as the short-term rate increases.     
\end{itemize}

This model can only have positive solutions since when the interest rate is $0$ it ends up being positive later on. Also, when it is low or close to 0, the standard deviation is close to 0.

\begin{figure}[ht]
    \centerline{\scalebox{0.3}{
    \includegraphics{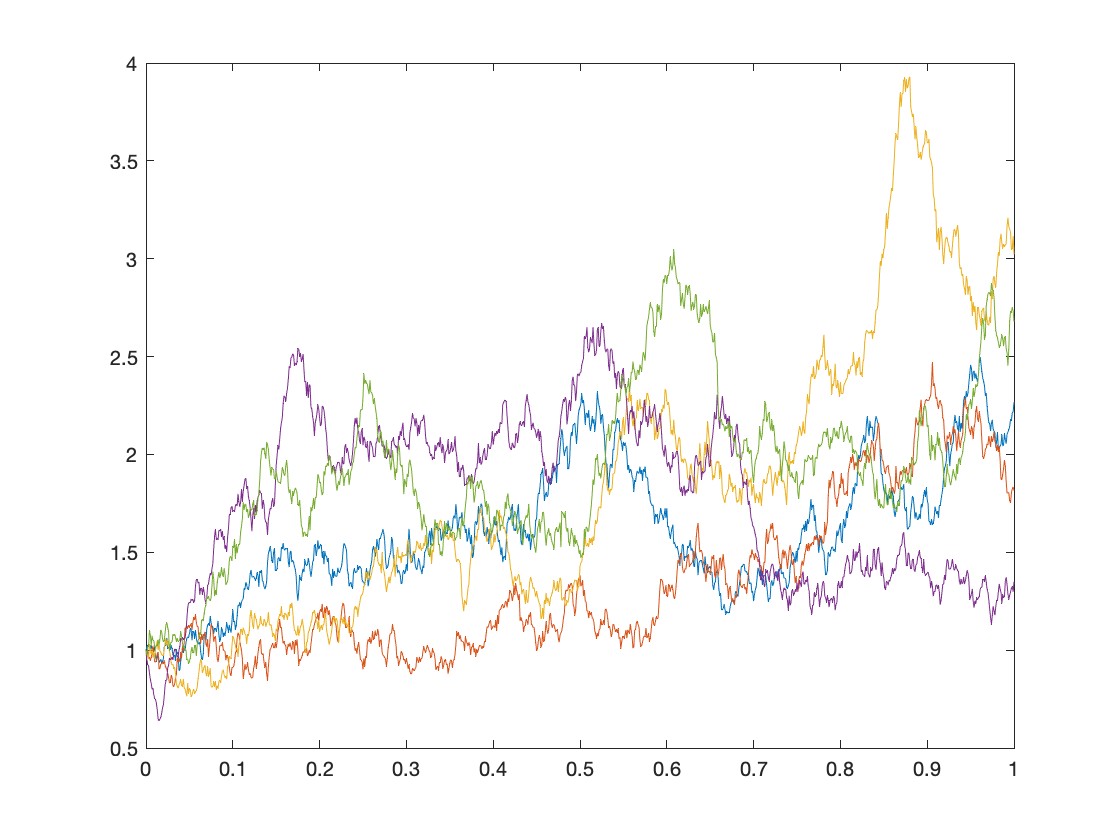}}}
    \caption{Simulated paths of the CIR model}
    \label{fig:22}
\end{figure}

The only solution to \eqref{equ:6} is what is known as the CIR process. Integrating Eq.  (\ref{equ:6}):
\begin{equation}
X_t=X_s+\alpha\int_s^t (\mu - X_u)du+\sigma \int_s^t\sqrt{X_u}dW_u, \hspace{1cm} s<t
\end{equation}
therefore

$$E[X_t|X_s]=X_s+\alpha\int_s^t (\mu-E[X_t|X_s])du, \hspace{1cm} s<t$$

If we call $m_t=E[X_t|X_s]$ we have

$$\frac{d}{dt}m_t=\alpha (\mu-m_t),\hspace{1cm} s<t$$

whose solution is:

$$m_t=X_se^{-\alpha(t-s)}+\mu(1-e^{-\alpha(t-s)})$$

So
\begin{equation}
E[X_t|X_s]=X_se^{-\alpha(t-s)}+\mu(1-e^{-\alpha(t-s)}),\hspace{1cm} s<t
\end{equation}
and therefore
\begin{equation}
E[X_t|X_s]-\mu=(X_s-\mu)e^{-\alpha(t-s)},\hspace{1cm} s<t \label{SolutionExpectedCIR}
\end{equation}
Thus $E[X_t|X_s]-\mu$ has the same sign as $X_s-\mu$. In addition, if $\mu>0$ and $\alpha>0$, starting with $X_s>0$ we conclude that $X_t>0$.

\begin{figure}[ht]
    \centerline{\scalebox{0.3}{
    \includegraphics{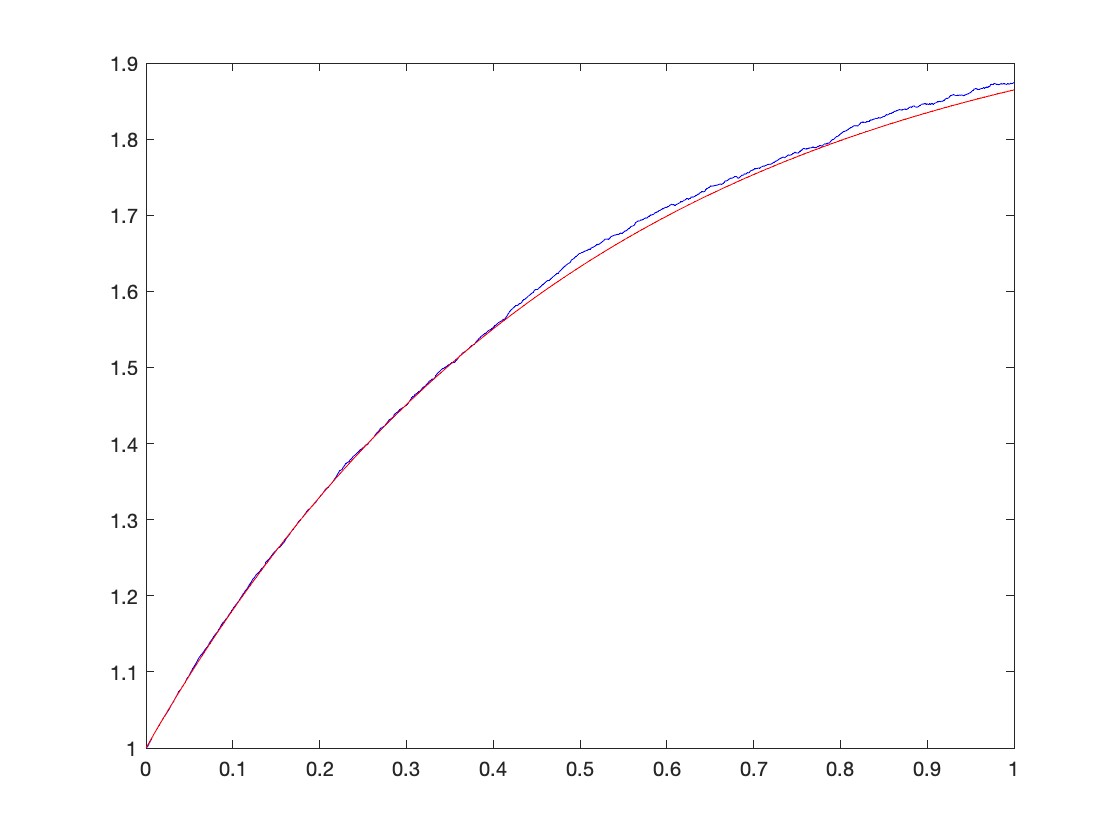}
    }}
    \caption{    
    Numerical (blue) and theoretical (red) mean comparison of the CIR model with Euler's method for $1000$ paths, with $x_0=1$, $\alpha= 2$, $\mu=2$ and $\sigma=1 $ and $5000$ subintervals.}
    \label{fig:20}
\end{figure}

Similarly, the variance is found to be:
\begin{equation}
Var[X_t|X_s]=\frac{X_s\sigma^2}{\alpha}(e^{-\alpha(t-s)}-e^{-2\alpha(t-s)})+\frac{\mu \sigma^2}{2\alpha}(1-e^{-\alpha(t-s)})^2
\end{equation}

\begin{figure}[!ht]
    \centerline{\scalebox{0.3}{
    \includegraphics{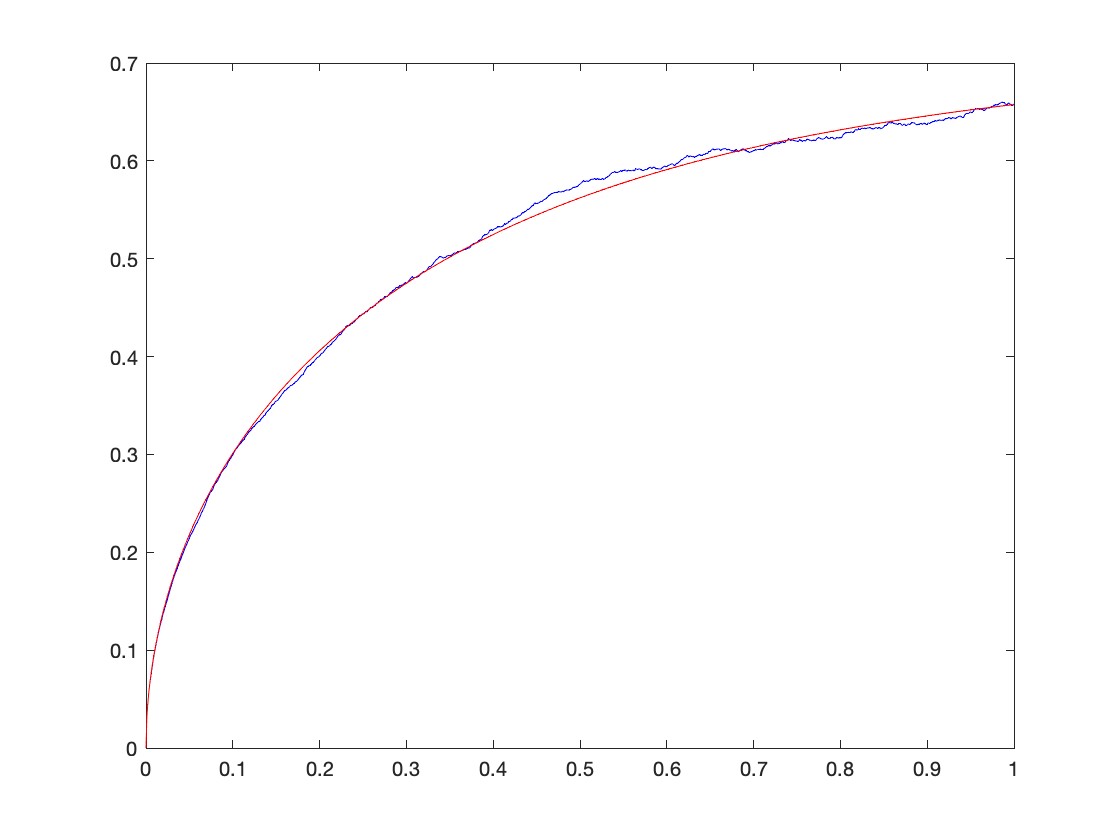}
    }}
    \caption{Comparison of the numerical (blue) and theoretical (red) standard deviation of the CIR model with the Euler method for paths of $1000$, with $x_0=1$, $\alpha= 2$, $\mu=2$ and $\sigma=1$ and 5,000 subintervals}
    \label{fig:21}
\end{figure}

As stated at the beginning, the fundamental advantage of this model is that the solutions are nonnegative. However, the distribution of the CIR model is not Gaussian, which makes it difficult to analyze.

The density function is given by:

$$f(X_s,s,X_t,t)=ce^{-(u+v)}\left (\frac{v}{u}\right)^\frac{q}{2}I_q(2\sqrt{uv})$$
{\setlength{\parindent}{0pt} where }
\begin{align*}
    c &=\frac{2\alpha}{\sigma^2(1-e^{-\alpha \Delta t})}\\
    u &=cX_se^{-\alpha \Delta t}\\
    v &=cX_t\\
    q &=\frac{2\alpha \mu}{\sigma^2}-1\\
    \Delta & t =t-s
\end{align*}
$I_q(\cdot)$ is a Bessel function of first type and order $q$:
\[
I_q(x)=\sum_{j=0}^{\infty}\left(\frac{x}{2}\right)^{2j+q}\frac{1}{k!\Gamma(j+q+1)}
\]
where $\Gamma$ is the gamma function.

Let $z_t=2cX_t$. Then the conditional distribution of $z_t$ given $z_s$ is an uncentered $\chi_d^2(2u)$, with $d=\frac{4\alpha\mu}{\sigma^2}$ degrees of freedom and the non-centrality parameter is $\lambda=2u$.

Therefore 
$$z_t|z_s \sim \chi^2(d,\lambda)$$
where:
\[
d=\frac{4\alpha\mu}{\sigma^2}
\]
\[
\lambda=\frac{4\alpha}{\sigma^2(1-e^{-\alpha \Delta t})}e^{-\alpha \Delta t}X_s
\]

Since $z_t=2cX_t$, $X_t$ conditional on $X_s$ has the same distribution as $z_t/2c$ conditional on $z_s/2c$. So,

$$X_t|X_s \sim \frac{z_t}{2c}|\frac{z_t}{2c} \sim \frac{1}{2c} \chi^2(d,\lambda)$$

We are going to make a study of the different behaviours that the deterministic solution of the CIR model equation can present in terms of the relations among the different parameters, which will be useful later to give an interpretation of the model parameters, although we know that the stochastic part would give oscillations with respect to said behaviour. We shall allow in this analysis for negative values of $\alpha$ since in certain regions of the data the calibrated values of $\alpha$ result in negative values.
%
%
%
%
According to Eq. (\ref{SolutionExpectedCIR})
we distinguish two cases, depending on whether $X_t$ is greater or less than $\mu$ and in each case two subcases, depending on whether $\alpha$ is positive or negative:

\begin{enumerate}
\item If $X_t<\mu$:
    \begin{itemize}
    \item If $\alpha>0$,  $X_t$ approaches $\mu$ from below.
         \item If $\alpha
         <0$,  $X_t$ moves away from $\mu$ downwards.
     \end{itemize}
    
\item If $X_t>\mu$:
    \begin{itemize}
    \item If $\alpha>0$, $X_t$ approaches $\mu$ from above.
         \item If $\alpha
         <0$, $X_t$ moves away from $\mu$ upwards.
    \end{itemize}
    
\end{enumerate}

\subsubsection{Estimation of the parameters}\label{EstimationParameters}

To approximate the data well, it is necessary to give a good adjustment of the parameters. In the case of the CIR* model, we must estimate three parameters, $\alpha$, $\mu$ and $\sigma$. We will generally refer to them as the parameter vector $\theta \equiv (\alpha,\mu,\sigma)$.
The procedure that will be followed to estimate the parameters is the one shown in \cite{kladivko2007maximum} and it is the maximum likelihood method (MLE), which is based on maximizing the objective function under consideration.
For the maximum likelihood estimation of the parameter vector $\theta \equiv (\alpha,\mu,\sigma)$ the transition densities are required. The CIR process is one of the processes for which we know its density function explicitly. Given $X_t$ at time $t$, the density of $X_{t+\Delta t}$ at time $t+\Delta t$ is:

$$p(X_{t+\Delta t}|X_t;\theta, \Delta t)=ce^{-(u+v)}\left (\frac{v}{u}\right)^\frac{q}{2}I_q(2\sqrt{uv})$$

{\setlength{\parindent}{0pt} where }
\begin{align*}
    c &=\frac{2\alpha}{\sigma^2(1-e^{-\alpha \Delta t})}\\
    u &=cX_te^{-\alpha \Delta t}\\
    v &=cX_{t+\Delta t}\\
    q &=\frac{2\alpha \mu}{\sigma^2}-1\\
\end{align*}

{\setlength{\parindent}{0pt}where $I_q(2\sqrt{uv})$ is a Bessel function.}

The likelihood function for time series with $N$ observations is:

\begin{equation}\label{eqn:20}
  L(\theta)= \prod_{i=1}^{N-1}p(X_{t_{i+1}}|X_{t_i};\theta, \Delta t)  
\end{equation}

To simplify the calculations, it is usual to work with the log-likelihood expression, which consists of taking logarithms in the equation \eqref{eqn:20}.

\begin{equation}
    \ln L(\theta)=\sum_{i=1}^{N-1}\ln p(X_{t_{i+1}}|X_{t_i};\theta, \Delta t)
\end{equation}

{\setlength{\parindent}{0pt}from which the log-likelihood function of the CIR process can be easily derived.}

\begin{equation}\label{eqn:21}
    \ln L(\theta)=(N-1)\ln c+\sum_{i=1}^{N-1}\left[ -u_{t_i}-v_{t_{i+1}}+\frac{1}{2}q\ln \left(\frac{v_{t_{i+1}}}{u_{t_i}}\right) + \ln\left(I_q\left(2\sqrt{u_{t_i}v_{t_{i+1}}}\right)\right) \right]
\end{equation}

{\setlength{\parindent}{0pt} where $u_{t_i}=cX_{t_i}e^{-\alpha\Delta t}$ y $v_{t_{i+1}}=cX_{t_{i+1}}$}

To find the maximum likelihood estimate $\hat{\theta}$ of the parameter vector $\theta$, we have to maximize the function \eqref{eqn:21} over its parameter space.

\begin{equation}\label{eqn:22}
    \hat{\theta}=(\hat{\alpha}, \hat{\mu}, \hat{\sigma})=\arg \underset{\theta}{\max} \ln L(\theta)
\end{equation}

Since the logarithm function is monotonically increasing, maximizing the log-likelihood function is equivalent to maximizing the likelihood function.

To solve the problem \eqref{eqn:22} we resort to numerical computation.
For the global optimal convergence, the initial optimization points are essential, for which the method of least squares will be used.
We first write the equation of the discretized CIR*:

\begin{equation}\label{eqn:23}
    X_{t+\Delta t}-X_t=\alpha (\mu-X_t)\Delta t + \sigma\sqrt{X_t} W_t
\end{equation}

{\setlength{\parindent}{0pt}where $W_t$ is distributed with zero mean and variance $\Delta t$.}

Dividing the equation \eqref{eqn:23} by $\sqrt{X_t}$ we get:

$$\frac{X_{t+\Delta t}-X_t}{\sqrt{X_t}}=\frac{\alpha\mu \Delta t}{\sqrt{X_t}}-\alpha \sqrt{X_t}\Delta t+\sigma W_t$$

Based on this expression, the initial values of $\hat{\alpha}$ and $\hat{\mu}$ are found by minimizing the function:

$$(\hat{\alpha},\hat{\mu})=\arg \underset{\alpha, \mu}{\min}\sum_{i=1}^{N-1}\left(\frac{X_{t_{i+1}}-X_{t_i}}{\sqrt{X_{t_i}}}-\frac{\alpha\mu \Delta t}{\sqrt{X_{t_i}}}+\alpha \sqrt{X_{t_i}}\Delta t\right)^2$$

The exact expression of the solution is found on page 3 of \cite{kladivko2007maximum}. The estimate of $\hat{\sigma}$ is found as the standard deviation of the residuals.

To optimize the objective function \eqref{eqn:21} we need to evaluate the Bessel function $I_q(2\sqrt{uv})$. The function \textit{besseli} is implemented in Matlab, but this usually causes problems, because the function $I_q=(2\sqrt{uv})$ approaches infinity very quickly. Fortunately, Matlab allows us to give another scaled version, which we will call $I_q^1(2\sqrt{uv})$, which solves the divergence problem in such a way that:

$$I_q^1(2\sqrt{uv})=I_q(2\sqrt{uv})\exp(-2\sqrt{uv})$$

And therefore:
$$I_q(2\sqrt{uv})=\frac{I_q^1(2\sqrt{uv})}{\exp(-2\sqrt{uv})}$$

Rewriting the expression \eqref{eqn:21} we get:

\begin{align*}
\ln L(\theta)=(N-1)\ln c &+\sum_{i=1}^{N-1}
(-u_{t_i}-v_{t_{i+1}}+\frac{1}{2}\ln \left(\frac{v_{t_{i+1}}}{u_{t_i}}\right)+\\
&+ \ln\left(I_q^1(2\sqrt{u_{t_i}v_{t_{i+1}}})\right)+2\sqrt{u_{t_i}v_{t_{i+1}}} )
\end{align*}

\subsubsection{Numerical methods}

To obtain an approximation of the exact solution of the equation we need to establish a numerical scheme. In our case, we will implement the Euler-Maruyama and Milstein numerical schemes, to see later if there are notable differences between them.

The Euler-Maruyama scheme or method is an extension of Euler's method for ordinary differential equations to stochastic differential equations.
Let be an Itô process $\{X_t,0 \leq t \leq T\}$ that is the solution of the following stochastic differential equation:

\begin{equation}\label{equ:11}
\left\lbrace
\begin{array}{ll}
dX_t=f(t,X_t)dt+ g(t,X_t)dW_t\\
X_0=x_0
\end{array}
\right.
\end{equation}

{\setlength{\parindent}{0pt}where $W(t)$ represents the Wiener process and suppose we want to solve this SDE in the time interval $[0,T]$.}

The Euler-Maruyama approximation $Y_i$ to the true solution of $X$ is defined as follows:
\begin{itemize}
    \item Divide the interval $[0,T]$ in $N$ subinterval of size $\Delta t= T/N$ being $0=t_0<t_1<...<t_N=T$.
    \item Set the initial condition $Y_0=x_0$.
    \item Define recursively $Y_i$ for $1\leq i\leq N$
    
    \begin{equation}\label{equ:10}
    Y_{i+1}=Y_i+f(t_i,Y_i)\Delta t+g(t_i, Y_i)\Delta W_i
\end{equation}

where $\Delta W_i=W_{t_i+1}-W_{t_i}$.
\end{itemize}

The variables $\Delta W_i$ are independent and identically distributed normal random variables, that is, $\Delta W_i \sim \sqrt{\Delta t}Z$ with $Z \sim N(0,1)$.

The Milstein method \cite{mil1975approximate} is used to increase the accuracy of the Euler-Maruyama method. This is achieved by introducing a term of order 2 by using the partial derivative with respect to $x$ of $g(t,x)$.

Given an Itô process $\{X_t,0 \leq t \leq T\}$ which is a solution of the stochastic differential equation \eqref{equ:11} the approximation of the Milstein method $Y_i$ to the true solution of $ X$ is given by:

\begin{itemize}
    \item Divide the interval $[0,T]$ into subintervals of size $\Delta t=T/N$ with $0=t_0 < t_1< ...< t_n=N$.
     \item Take as initial condition $Y_0=x_0$.
     \item Recursively define $Y_i$ for $1\leq i\leq N$ by:    
    \begin{equation}\label{equ:13}
     Y_{i+1}=Y_i+f(t_i,Y_i)\Delta t+g(t_i, Y_i)\Delta W_i+\frac{1}{2}g(t_i, Y_i)\frac{\partial g(t_i, Y_i)}{\partial x}[(\Delta W_i)^2-\Delta t]
    \end{equation}
  where $\Delta W_i=W_{t_{i+1}}-W_{t_i}$.  
    
\end{itemize}

The variables $\Delta W_i$ are independent and identically distributed normal random variables, that is, $\Delta W_i \sim \sqrt{\Delta t}Z$ with $Z \sim N(0,1)$.
As we can see, the expression of the scheme is the same as that of Euler, except that the summand is added

$$\frac{1}{2}g(t_i, Y_i)\frac{\partial g(t_i, Y_i)}{\partial x}[(\Delta W_i)^2-\Delta t]$$

Therefore, if $\frac{\partial g(t,x)}{\partial x}$ turns out to be $0$ this method is equivalent to the Euler-Maruyama method.
When a method satisfies $E\left(|Y_i-X(t_i)|\right) \leq K (\Delta t)^\gamma$ for some $\gamma$, that method is said to be a strong approximation of order $\gamma$.
Applying this, the Euler-Maruyama method is a strong approximation of order $\gamma=1/2$ while the Milstein method is a strong approximation of order $\gamma=1$ if $f(t, X_t)$ and $g(t,X_t)$ are $\mathcal{C}^1$ functions.
The functions with which we are working in these models comply with this, so the order of convergence of the Milstein method will always be greater than that of Euler-Maruyama.

\section{Results} \label{Result}

Since we have daily data we set the time step $\Delta t=1$. To compare both the Euler-Maruyama and the Milstein methods and see that they fit the real data we have well, we are going to choose a specific time. A window of 100 data will be taken from the first real data available and the following 500 days will be estimated. The prediction is made for the day following the last one of the windows, the window is moved one unit to the right and the process is repeated. 

\begin{figure}[!ht]
    \centerline{\scalebox{0.3}{
    \includegraphics{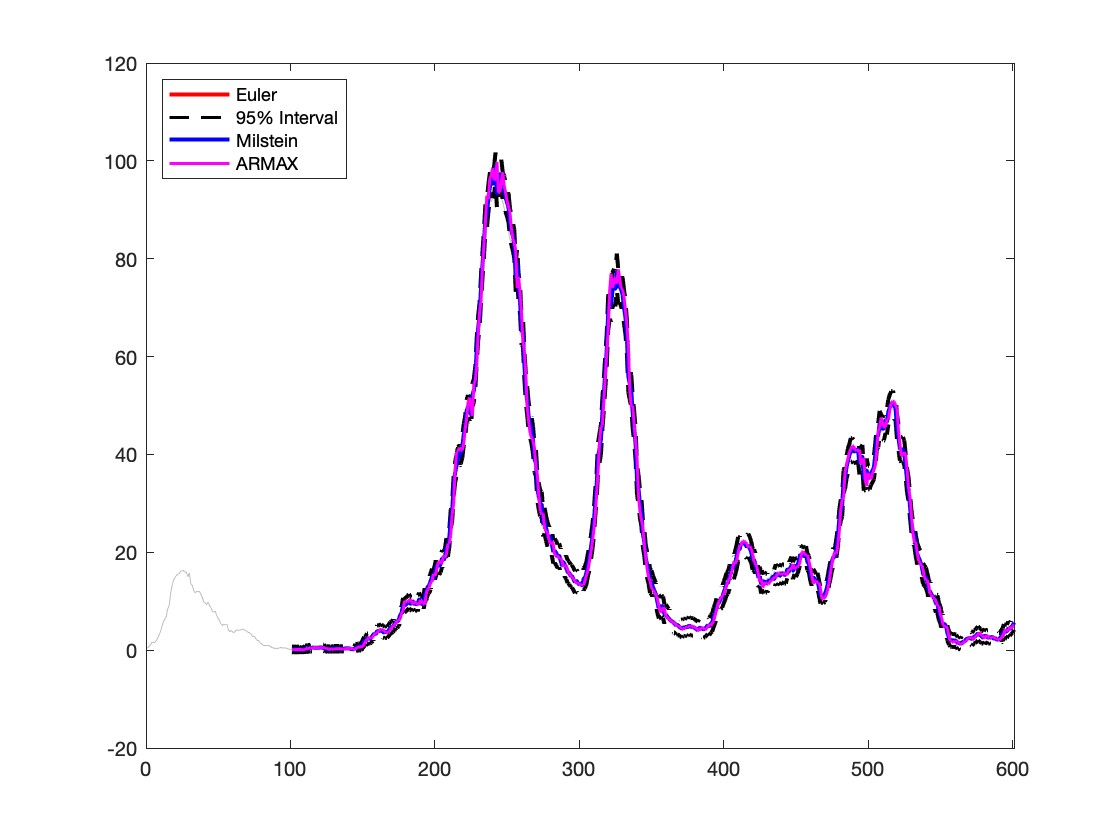}
    }}
    \caption{Approximations of the CIR*  with Euler-Maruyama and Milstein, and ARMAX with win=100, tin=1 and nsim=1000.}
    \label{fig:fig20}
\end{figure}

\begin{figure}[!ht]
    \centerline{\scalebox{0.2}{
    \includegraphics{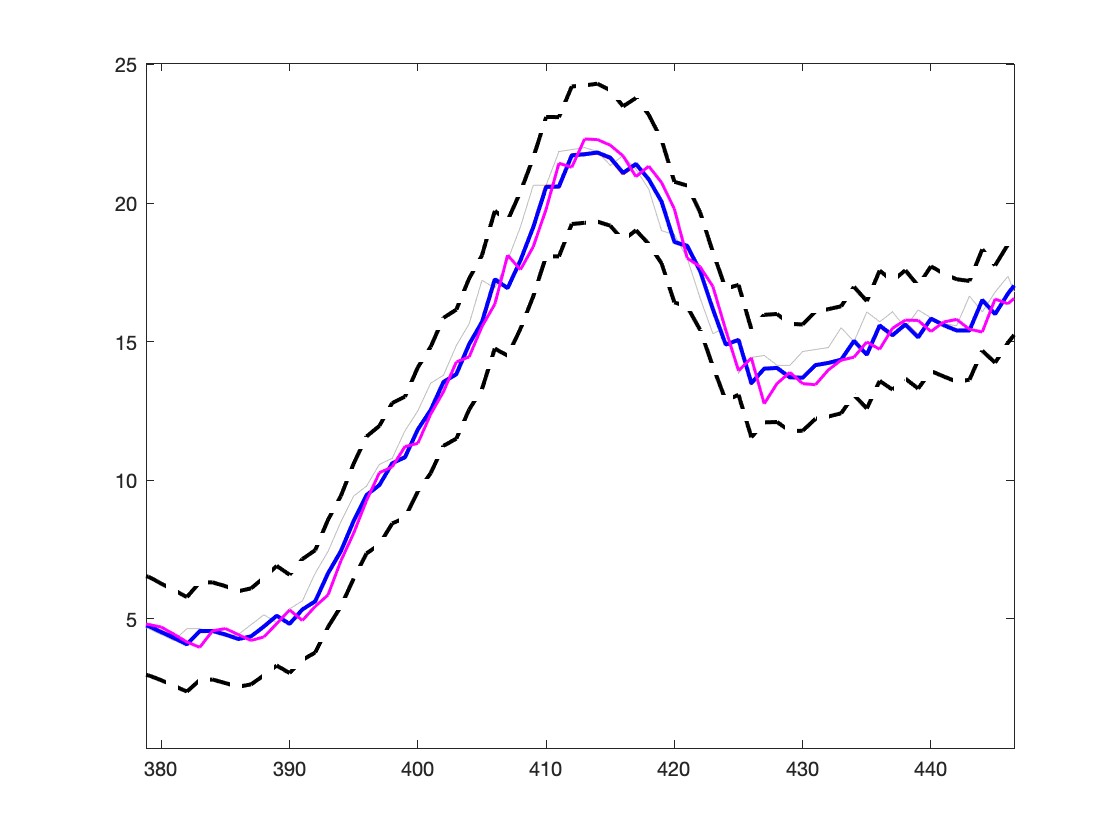}
    \includegraphics{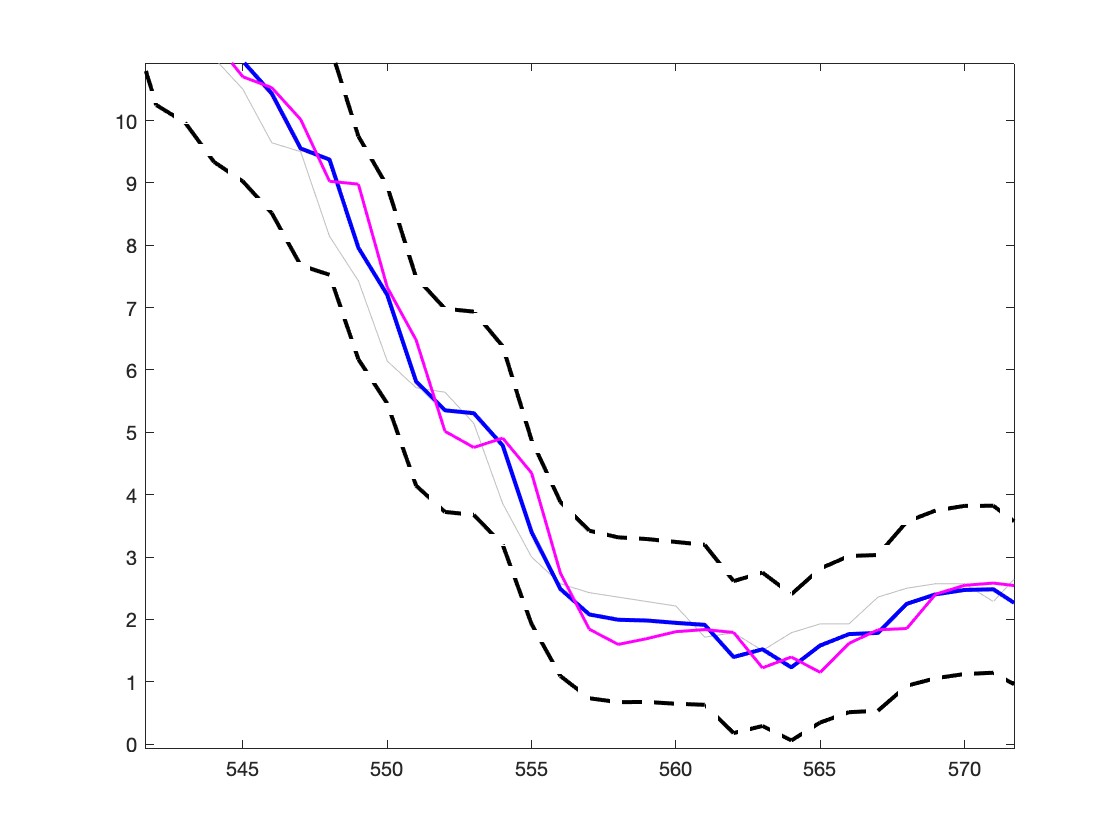}}}
    \caption{Comparison of CIR* (blue) and ARMAX models (magenta) }
    \label{fig:figzoom}
\end{figure}

Figure \ref{fig:fig20} shows the three estimates that we want to compare, but as is logical, with that size they are not seen in detail. This is because the figure shows a first glance at the obtained estimates.
Next, Figure \ref{fig:figzoom} represents the real data (in grey), the ARMAX model estimates (in magenta) and the ones calculated with the CIR* model (blue curve). As in the previous section, the Euler-Maruyama and Milstein methods are very similar, so the differences between the depicted curves cannot be seen unless a larger zoom is made.

It can be seen that the blue curve generally fits the real data better, so it stands to reason that the CIR* model is slightly better than the ARMAX model. To verify this rigorously, the root mean square error (RMSE) of the CIR* model has been calculated with the Euler-Maruyama and Milstein method, and the RMSE of the ARMAX model. These errors are collected in the following Table \ref{table:comp}. As we can see, the CIR* model gives better results than the ARMAX model.

\begin{table}[!ht]
\centering
\begin{tabular}{l|l|l|l|}
\cline{2-4}
                          & CIR* with Euler               & CIR* with Milstein            & ARMAX                       \\ \hline
\multicolumn{1}{|l|}{RMSE} & \multicolumn{1}{c|}{1.0556} & \multicolumn{1}{c|}{1.0558} & \multicolumn{1}{c|}{2.2157} \\ \hline
\end{tabular}
\caption{Comparison of CIR* model errors (Euler-Maruyama and Milstein) and ARMAX}
\label{table:comp}
\end{table}

Finally, we are going to give an interpretation of the parameters of the CIR* model at different stages of the pandemic, based on the analysis of the deterministic solution of the CIR* model given just before  Sec. \ref{EstimationParameters}.

We begin by studying a stage in which infections are increasing. 
For the estimated parameters to have the same trend, the window from which they are estimated must also be in the growth range, which will force us to take a small window and estimate few values, since if we look at Figure \ref{fig:23}, we see that the periods in which the infections grow are not many days. Taking this into account, we are going to focus on the section that goes from day 210 to 229, that is, 20 forecasts, taking a window of 30 days.
Based on the results obtained, we observe that both the mean of $\alpha$ and $\mu$ are negative. Since the estimated data values are greater than the mean, then $X_t$ would move away from the mean upwards.

\begin{figure}[!ht]
    \centerline{\scalebox{0.2}{
    \includegraphics{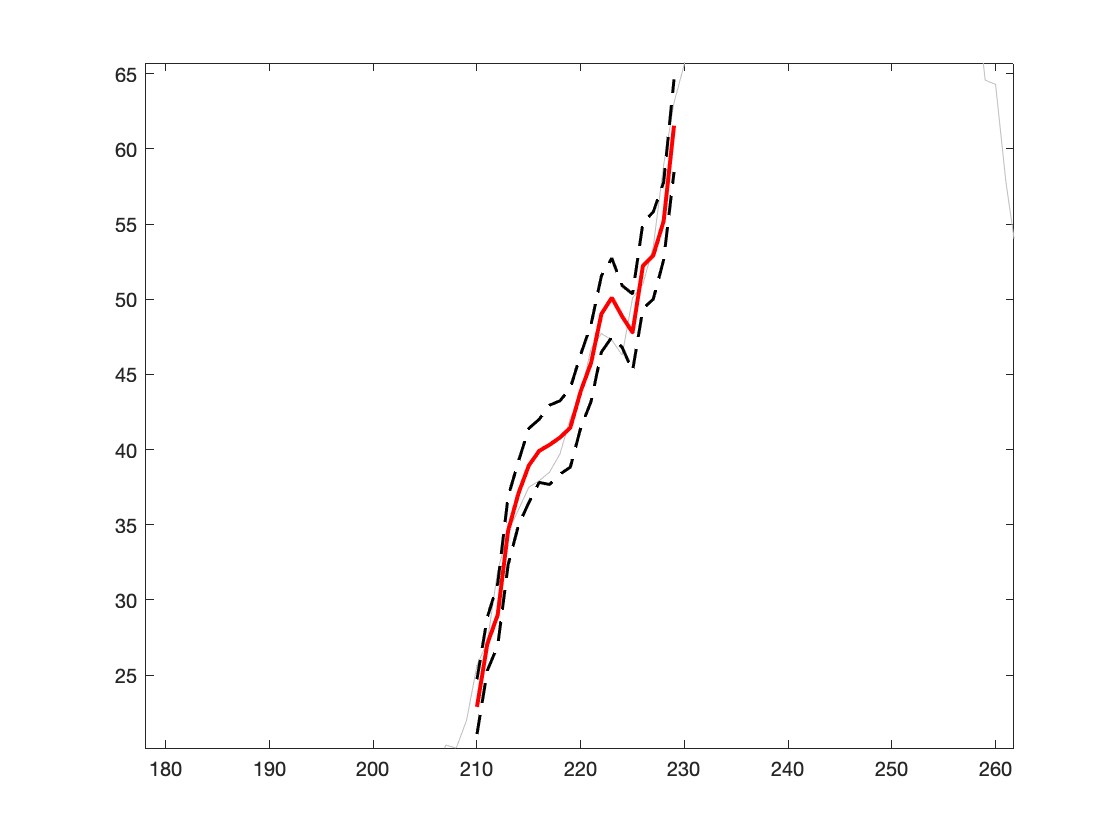}
    \includegraphics{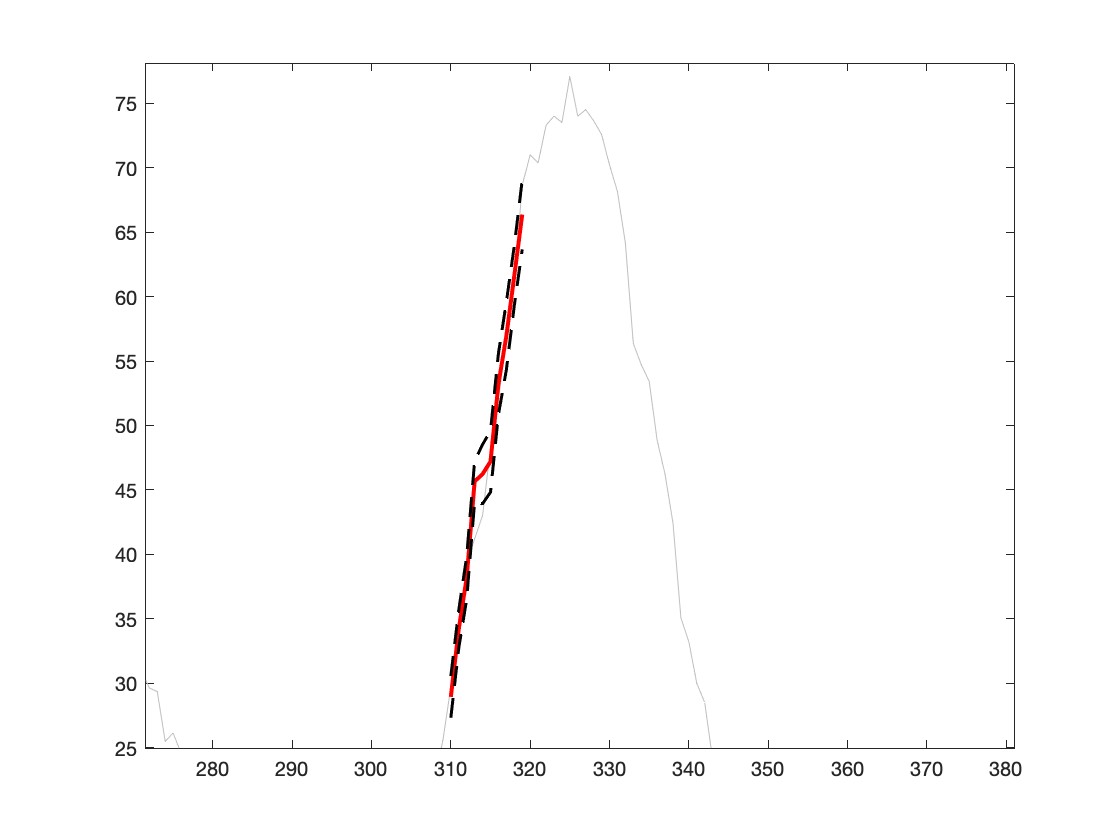}}}
    \caption{Stage 1 (left) and stage 2 (right) of a steep growth of the COVID-19 pandemic daily cases.}
    \label{fig:figcreciente}
\end{figure}

To corroborate that this is true, another stage of increase in cases of the pandemic has been taken, specifically from day 310 to 319, that is, 10 days, and a window of 30 days. The number of forecasts had to be reduced because, as previously mentioned, the window of days must be in the growth range for good analysis. 
In Figure \ref{fig:figcreciente} one can see both sections and in Table \ref{tabla:2}, the exact values of the mean of the parameters in the two stages.

\begin{table}[!ht]
\centering
\begin{tabular}{l|c|c|}
\cline{2-3}
                            & Stage 1 & Stage 2 \\ \hline
\multicolumn{1}{|l|}{Average $\alpha$} & -0.0562   & -0.0990 \\ \hline
\multicolumn{1}{|l|}{Average $\mu$} & -3.6346   & -3.5105 \\ \hline
\multicolumn{1}{|l|}{Average $\sigma$} & 0.2253    & 0.2007  \\ \hline
\end{tabular}
\caption{Growing stages of COVID-19 infections}
\label{tabla:2}
\end{table}

Keeping the ideas we used from the previous case, we now take a time in which the cases decrease. We estimate the time step that goes from day 265 to 284. In this case, we again obtain the negative mean of $\alpha$, which makes sense, since in periods of strong growth or decrease the values move away from the mean. On the contrary, now the mean of the parameter $\mu$ is very large and exceeds the mean of infections in that section, therefore, the values $X$ are far from the mean, but this time below.

\begin{figure}[!ht]
    \centerline{\scalebox{0.2}{
    \includegraphics{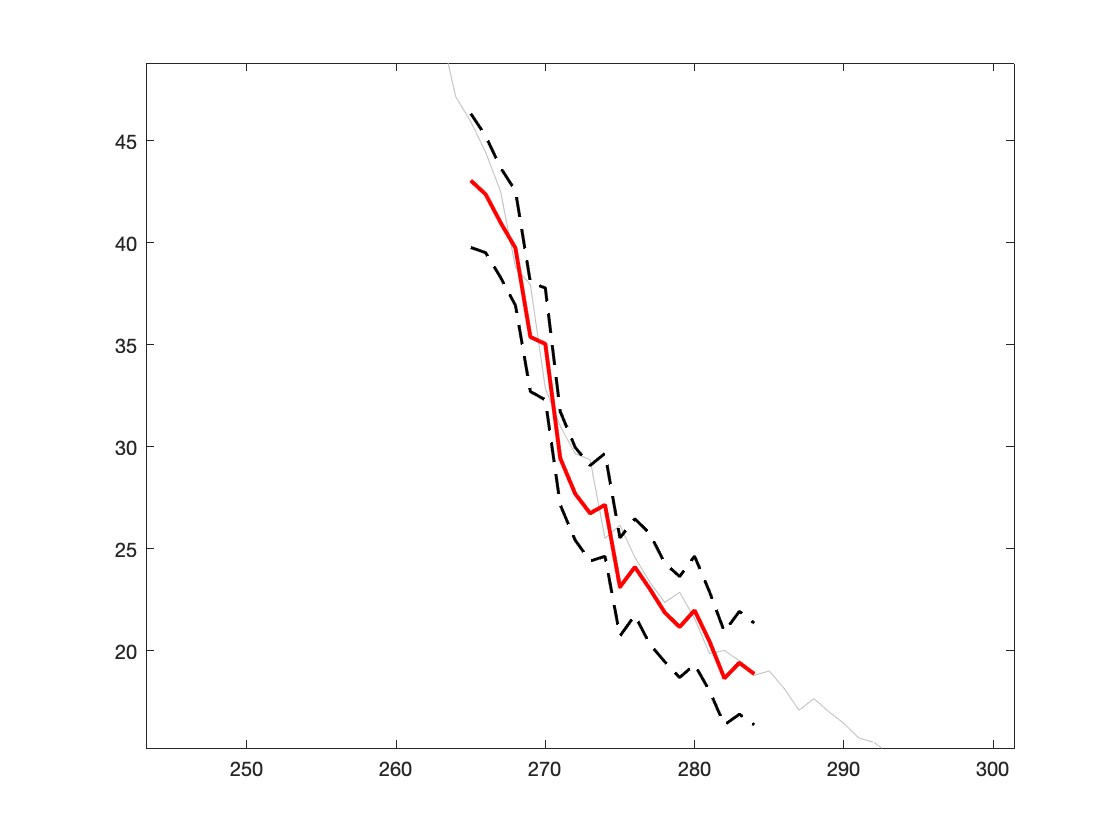}
    \includegraphics{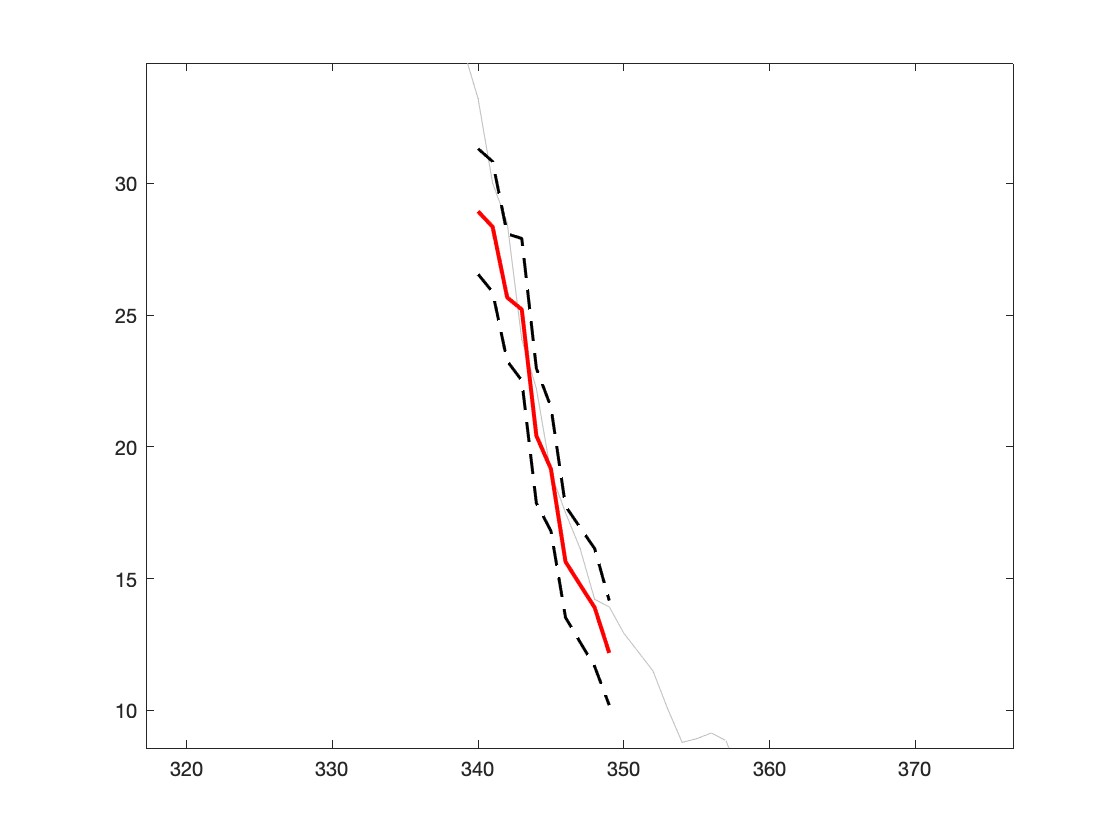}}}
    \caption{Stage 3 (left) and stage 4 (right) of a decisive decrease in the daily cases of the COVID-19 pandemic. }
    \label{fig:figdecreasing}
\end{figure}

As in the previous case, another section has been taken to verify the results. The estimates of both stages can be seen in Figure \ref{fig:figdecreasing} and the comparison of the mean values of the parameters in Table \ref{table:3}.

\begin{table}[!ht]
\centering
\begin{tabular}{l|c|c|}
\cline{2-3}
                            & Stage 3 & Stage 4 \\ \hline
\multicolumn{1}{|l|}{Average $\alpha$} & -0.01664   & -0.0305 \\ \hline
\multicolumn{1}{|l|}{Average $\mu$} & 190.5401   & 95.5082 \\ \hline
\multicolumn{1}{|l|}{Average $\sigma$} & 0.2644    & 0.2925  \\ \hline
\end{tabular}
\caption{Decreasing of COVID-19 infections}
\label{table:3}
\end{table}

Finally, we are going to interpret the parameters in a section where the values are relatively constant.

As seen in figure \ref{fig:23}, there are a few sections where this occurs. We are going to take days between 110 and 139, in which it is observed that the COVID cases are close to the 0 value. This may seem surprising, but it makes sense because it precisely coincides with the state of alarm. As can be seen on the right side of the graph, from day 150 the cases begin to increase, coinciding with the de-escalation process and the summer of 2020 when restrictions were relaxed.

\begin{figure}[!ht]
    \centerline{\scalebox{0.3}{
    \includegraphics{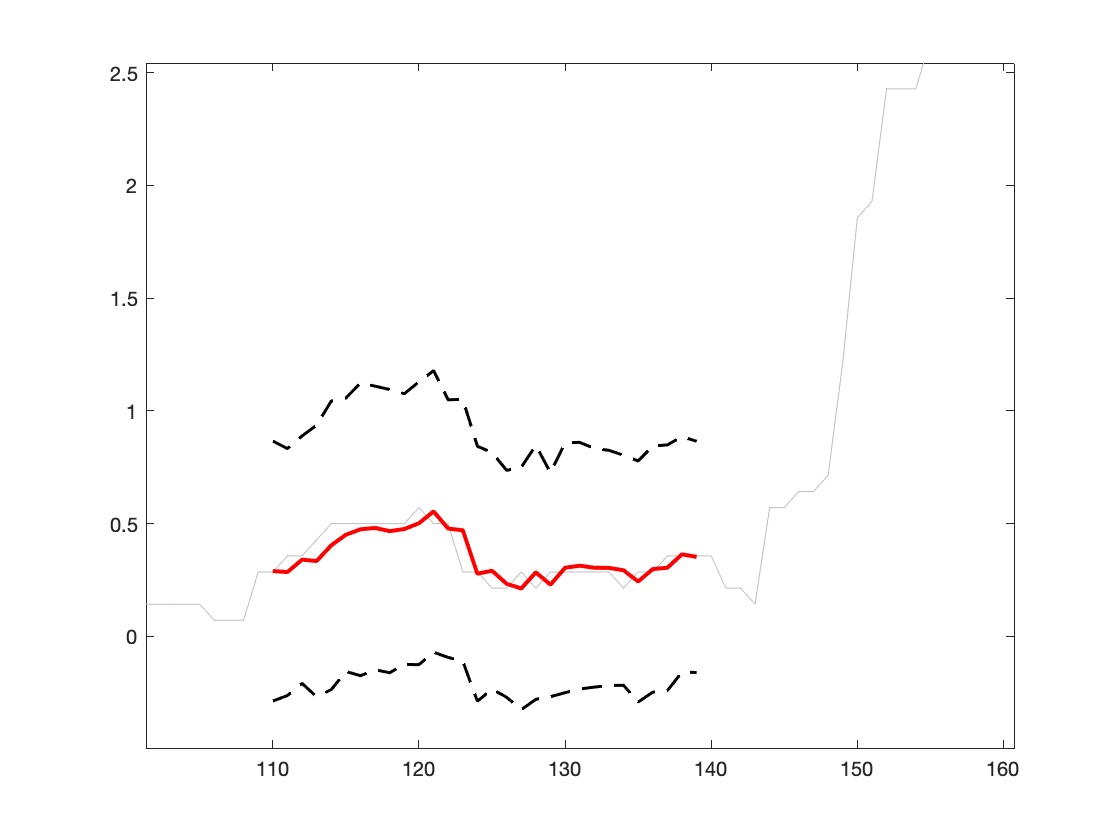}
    }}
    \caption{Example of a stage of the COVID-19 pandemic where the number of cases are relatively flat.}
    \label{fig:fig35}
\end{figure}

At this stage, the mean of the parameters $\alpha$ is 0.1397, that is, positive, unlike the previous cases. The mean of the parameters $\mu$ is 0.3303 and that of $\sigma$ is 0.1448, so we have a small deviation.

\begin{table}[!ht]
\centering
\begin{tabular}{l|c|}
\cline{2-2}
                            & Stage 5 \\ \hline
\multicolumn{1}{|l|}{Average $\alpha$} & 0.1397   \\ \hline
\multicolumn{1}{|l|}{Average $\mu$} & 0.3303   \\ \hline
\multicolumn{1}{|l|}{Average $\sigma$} & 0.1448    \\ \hline
\end{tabular}
\caption{Relatively constant stage of COVID-19 infections.}
\label{tabla:4}
\end{table}

According to the analysis of the deterministic solution, $X_t$ will oscillate around the mean, which makes sense, since the real mean of those days is around $0.33$ and the value of the data varies between  $0.25$ and $0.55$ cases.

\section{Conclusions} \label{Conclusion}
The main objective of this work is the study and development of some stochastic models typically used in financial markets applied to the COVID-19 pandemic in the city of Jaén.

For solving stochastic equations both the Euler-Maruyama method and the Milstein method were used with reference to the CIR* stochastic process. Over the reported COVID-19 daily cases of the pandemic in the city of Jaén (Spain), the maximum likelihood method was used for parameters calibration. The forecasts given with this model have been validated with 95\% confidence intervals and with statistical measures of goodness of fit, such as the RMSE (root mean square error). The results obtained are consistent with the observations and quite accurate. For comparison, the classical ARIMAX model has been used, resulting in more accurate predictions for the suggested CIR* model. The reason could be the relatively small size of the city of Jaén, causing large fluctuations in the number of cases that are not sufficiently softened by the moving averages, resulting in a worse behaviour of ARIMAX in comparison with CIR*. The importance of the suggested approach is twofold because it not only extends the models available to scholars to model pandemics, but also paves the way for similar approaches in which financial models can be converted into econometric models.

Future research could be aimed at enlarging the scope to all provinces of Andalusia. In such a case, we could expect that a greater number of data could imply a longer time for the trend to change. In addition, the number of healed and deceased could also be studied, although the latter, being much smaller, will present the aforementioned problems. In terms of considered models, future research could include a comparison with the more advanced CIR\# by Orlando et al. \cite{Orlando2019, orlando2018new, orlando2020forecasting, Orlando2021Dec}.
In addition, although our work has only been done for one equation, it could also be generalized to systems of equations to discover the interrelation between different cities.
Finally, note that stochastic differential equations are not only a very powerful tool for modelling economic-financial variables, but also in the epidemiological field, being proven from this practical point of view.

\section*{Acknowledgments}

G.O. is a member of the research group of GNAMPA - INdAM (Italy).
J.G. acknowledges  Spanish MICINN  through the project PGC2018-097831-B-I00 and  Junta de Andaluc\'\i a through the project  FEDER-UJA-1381026.

\section*{Conflict of interest}

The authors declare there is no conflict of interest.

\bibliographystyle{vancouver} 
\bibliography{MyBibCovidJaen}


\end{document}